\documentclass[%
 reprint,
 amsmath,amssymb,
 aps,
 prb,
 floatfix,
 showkeys
]{revtex4-2}

\usepackage{graphicx}
\usepackage{dcolumn}
\usepackage{amsmath}
\usepackage{siunitx}
\usepackage{bm}
\usepackage{bm}
\usepackage{hyperref}
\usepackage{float}
\usepackage[
    type={CC},
    modifier={by},
    version={4.0},
]{doclicense}
\usepackage{orcidlink}
\usepackage{natbib} 

\usepackage[final]{./spin-textures}

\begin{document}
\doclicenseThis
\preprint{APS/123-QED}

\title{Simulation of current-driven magnetisation switching in nanopillars with Perpendicular Shape Anisotropy}

\author{Natalia Boscolo Meneguolo\textsuperscript{1}\,\orcidlink{0009-0000-7877-6819}}
 \email{natalia.boscolomeneguolo@cea.fr}%
 \author{Olivier Fruchart\textsuperscript{1}\,\orcidlink{0000-0001-7717-5229}}%
 \author{Jean-Christophe Toussaint\textsuperscript{2}\,\orcidlink{0000-0001-5382-3594}}%
 \author{Mouad Fattouhi\textsuperscript{1}\,\orcidlink{0000-0002-3366-6461}}%
 \author{Liliana D. Buda-Prejbeanu\textsuperscript{1}\,\orcidlink{https://orcid.org/0000-0002-6105-151X}}%
\author{Ioan-Lucian Prejbeanu\textsuperscript{1}\,\orcidlink{0000-0001-6577-032X}}%

\author{Daria Gusakova\textsuperscript{1}\,\orcidlink{0000-0002-5111-7079}}%

\affiliation{%
 \textsuperscript{1}Univ. Grenoble Alpes, CEA, CNRS, Grenoble INP, SPINTEC, 38000 Grenoble, France
}%
\affiliation{%
 \textsuperscript{2}Univ. Grenoble Alpes, CNRS, Institut Néel, 38000 Grenoble, France
}%


\begin{abstract}
The Perpendicular Shape Anisotropy Spin Transfer Torque Magnetic Random Access Memory (PSA-STT-MRAM) is a recent concept proposed to maintain the thermal stability of standard MRAM at small diameters, considering thick vertical pillars as the free layer. In order to explore the specific physics of \olivierReplace{PSA-STT-MRAMs expected in relation with their}{such systems with a} three-dimensional nature, we have performed simulations combining a micromagnetic model coupled self-consistently with spin-dependent transport equations. The 3D shape induces flower states at the upper and lower surfaces. Besides, the field-like component of STT is found to be larger than in \olivierReplace{standard MRAMs}{thin elements such a in standard MRAMs}, suggesting that it needs to be considered. The combination of both effects leads to the efficient excitation of edge ferromagnetic resonance modes, playing a key role in magnetisation reversal. These results highlight features of 3D nanomagnetic systems, largely disregarded so far\olivierReplace{, which need to be considered to optimise PSA-STT-MRAM to be a competitive solution for technological
implementation}{. These must be considered to reliably describe PSA-STT-MRAM, but more broadly in the rising field of 3D nanomagnetism}.
\end{abstract}

\keywords{Micromagnetism, Finite Element Method, Simulations, 3D nanomagnetic system, spin accumulation, STT, MRAM}
\maketitle


\section{\label{sec:1level1}Introduction}
Over the recent years the interest in 3D nanomagnetism has steadily grown, with both fundamental and applied interests \cite{FER2017}. On one side 3D nano-objects allow the existence and the tailoring of new magnetic properties related to curvature and finite size: effective anisotropy and Dzyaloshinskii–Moriya interaction, static or dynamic chiral effects etc. On the other side, these may find practical application that take advantage of the effects specific to 3D nanomagnetism \cite{STR2021}: increased stability against thermal excitation, nano-oscillators with tunable frequency, flexible nano-membranes with tunable magnetic properties, magnetic interconnects able to merge memory and logic units \cite{ONO1999,CHU2015} etc. The third dimension also opens new prospects for spintronics \cite{RAF2022}. An example is networks of magnetic nanowires\cite{BUR2020,MEN2021}, which may be advantageous when thinking about 3D integrated magnetic devices for neuromorphic applications and in-memory computation, thanks to their enhanced connectivity.\\

The Perpendicular-Shape-Anisotropy Spin-Transfer-Torque Magnetic Random Access Memory (PSA-STT-MRAM) is another example making use of 3D concepts for applied spintronics \cite{PER2018,WAT2018}. The main objective of this concept for MRAMs is to boost the thermal stability of the free layer of the Magnetic Tunnel Junction (MTJ, the unit cell of the MRAM) at diameters down to a few nanometers, providing a scalable solution to the foreseen bottleneck of standard perpendicular MTJs below typically \qty{25}{\nano\meter}. This is achieved thanks to the increase in thickness of the MTJ free layer, taking the form of a vertical nanopillar, therefore, a 3D nanomagnetic system. Unfortunately, some drawbacks stem from the increase of the free layer thickness. We withstand namely to the high writing critical voltage and slow writing time. Indeed, in an MTJ the spin-transfer torque is  mostly an interfacial effect, therefore moderately efficient to switch the magnetisation of a 3D system. To address this problem, slicing the free-layer pillar into a multi-layer stack has been proposed \cite{IGA2024}. However, while this solution indeed reduces the current required for writing, it relies on the standard fabrication route of deposition plus aggressive etching routes \cite{PER2018, WAT2018}, so that it suffers from large device-to-device distributions. Also, the required physical vapor deposition cannot be implemented in microelectronics vias, that would provide a natural route for high-density networks\cite{FER2017}. Therefore, simulation of current-driven magnetisation switching in nanopillars remains relevant to identify their dynamics and identify routes to improve it. Such simulations have already been conducted\cite{PER2018, CAC2021}, in which the so-called adiabatic component of spin-transfer torque was implemented as an interfacial term, such as is routinely done in conventional thin free layers of MRAM\cite{TIM2015}.

In this article we refine these approaches \olivierAdd{by evaluating in more details the underlying microscopic physics} , considering two ingredients relevant to properly describe a 3D spintronic-nanomagnetic system. First, we employ a Finite-Element Method micromagnetic solver, to faithfully describe the curved shape of a nanopillar\cite{feeLLGood}. Second, we evaluate explicitly and consider self-consistently the spin accumulation in the entire nano-pillar. This allows us to derive explicitly the spin accumulation in the entire 3D structure, as well as the noteworthy contribution of the field-like spin torque to the physics of the system. The structure of this article is the following: in section II we describe the methods, presenting the main aspects of the \texttt{feeLLGood} micromagnetic solver: the Finite Element Method used for the meshing procedure, the physics underlying the modeling of the magnetisation dynamics, focusing namely on the spin accumulation computation. In section III we report the main simulation results, starting from the specificities related to the mesh size, continuing with the magnetisation behaviour at the edges of the pillar, and finishing with spin accumulation and the link between reversal mode and magnitude of the injected current density. In section IV we draw conclusions and perspectives.
\section{\label{sec:2level1}Methods}

\subsection{\label{sec:2level2.1}Finite Element Method micromagnetic solver}

There are a number of ways to discretise the specimen in micromagnetic simulations, \eg, the Finite-Difference Method (FDM), the Finite-Volume Method (FVM) and the Finite-Element Method (FEM) \cite{MAT1997, FAL2000}. In the present work we use \texttt{FeeLLGood}, a finite-element 3D micromagnetic solver developed over the past two decades at Institut Néel and SPINTEC in Grenoble\cite{feeLLGood}. The FEM method is very flexible and suited to discretize complex and curved shapes. In practice, the specimen volume is meshed with GMSH, an open source 3D finite element mesh generator \cite{GEU2009}. We chose a mesh of unstructured type, meaning that it has not a uniform pattern. More precisely, the discretization is obtained by first meshing the surface through the boundary recovery algorithm \cite{SI2010}, while the mesh points inside the volume are added through a 3D version of the Delaunay algorithm \cite{REB1993}.

\subsection{\label{sec:2level2.2}Zhang-Li-like STT}

The description of the dynamics of the system results from solving the Landau-Liftshits-Gilbert (LLG) equation:

\begin{eqnarray}
\frac{\partial \vectm}{\partial t}=-\mu_{0}\gamma(\vectm\times\Heff)+\alpha\left({\vectm\times\frac{\partial \vectm}{\partial t}}\right)
\label{eq:LLG}.
\end{eqnarray}
In equation (\ref{eq:LLG}) $\gamma > 0$ is the gyromagnetic ratio, $\mu_{0}$ is the vacuum magnetic permeability, $\vectm$ the unitary magnetisation (the value of the magnetisation normalized by the spontaneous magnetisation $\Ms$), $\Heff$ is the effective field and $\alpha$ is the Gilbert damping factor. More information can be found in Ref.\cite{SZA_Nov_2008, KRI2014, ALO2014}.
In addition to the LLG equation (\ref{eq:LLG}), we consider also the impact of the spin torques (STT) in the system by taking into account an extra term to (\ref{eq:LLG}): \olivierAdd{

\begin{eqnarray}
\vect T_{\mathrm{STT}}=\frac{1}{\tausd}\frac{\vects}{\Ms}\times\vectm
\label{eq:T_STT}
\end{eqnarray}
In equation (\ref{eq:T_STT}) $\tausd$ is the sd electrons interaction characteristic time}, while $\vects$ represents the spin accumulation. $T_{\STT}$ accounts for the mutual interaction between magnetisation and the spins of conduction electron.
Originally, the STT torque description has been simplified for a bilayer, exploiting the so-called LLGS equation (where S stands for Slonczewski). This corresponds to the formalism introduced for the first time by J. Slonczewski in 1996 \cite{SLO1996}:

\begin{eqnarray}
T_{\STT}=-\gamma~\mu_{0}~\nSTT~\frac{\hbar J}{2~ \lvert e\rvert}\frac{1}{\Ms t}~\vectm\times(\vectm\times\vectmpol)\nonumber \\
+\gamma~\mu_{0}~\nSTT^{'}~\frac{\hbar J}{2~\lvert e\rvert}\frac{1}{\Ms t}(\vectm\times\vectmpol)
\label{eq:LLGS}.
\end{eqnarray}
In equation (\ref{eq:LLGS}) $\nSTT$ represents the STT efficiency, whose value can be estimated from the tunnel magnetoresistance of the devices, $\hbar$ is the reduced Planck constant, $e$ is the electron charge, $\Ms$ is the value of the spontaneous magnetisation of the material, $J$ is the current density injected in the system, $t$ is the thickness of the magnetic free layer, $\vectm$ is the unit vector for magnetisation and $\vectmpol$ is the unit vector for magnetisation in the reference layer assumed to be fixed, separated by a tunnel barrier from the free layer considered. There are two spin torque terms in the right-hand side of \eqnref{eq:LLGS}. The first one is the so-called Damping-Like component (DL, also called in-plane), that drives magnetisation towards the local effective field. The second term is the Field Like component (FL, also called out-of-plane), which induces magnetisation to precess around the local effective magnetic field. Usually, in 2D systems the FL component is neglected, since it has been shown to not play a major role in the reversal mechanism \cite{TIM2015}.

Another conceptual way to describe the STT is to consider four contributions in total: two of them related to the change in time of magnetisation and the other two related to its variations in space. The time-related contributions can be accounted for by redefining the gyromagnetic ratio $\gamma$ and the Gilbert damping factor $\alpha$. The space-related terms instead, describe two main interactions: first, the ballistic motion and all the adiabatic processes of conduction electrons; second, the non-adiabatic spin torque that are consequences of the spatial mistracking of spins between conduction electrons and local magnetisation. This description is the Zhang-Li one, introduced in 2004 by S. Zhang and Z. Li \cite{ZHA2004}. \texttt{FeeLLGood} relies on this Zhang-Li-like description of the STT.\\
\olivierAdd{Nevertheless, we would like to highlight that we are considering a simplified model for spin-polarised tunneling effect. The purpose of the present report is indeed to outline the main physical phenomena taking place in nanomagnetic structures during the magnetisation reversal.
However, if willing to be more precise, other effects could be additionally considered \cite{BUT2001}. For example the type of tunnel barrier \cite{YUA2007} or the applied switching bias \cite{KAL2009} could affect the STT.}

\subsection{\label{sec:2level2.3}Spin accumulation}

Spin-transfer torques result from spin accumulation, \ie, an out-of-equilibrium imbalance of up and down spins of the conduction electrons. There can be different implementations of STTs. When dealing with standard MRAM with perpendicular magnetization, which have an ultrathin layer, one generally assumes that the spin accumulation is uniform across the free layer. Indeed, as already mentioned in the introduction, usually the effiency of the damping-like coupling can be computed from experimentally obtained current vs field switching diagrams \cite{TIM2015}.
However, this approach may be challenged when dealing with 3D structures. An ersatz consist in considering an effective interfacial STT, \eg, decaying away from the interface in an effective exponential way\cite{CAC2021}. However, this choice can also be questionable, not considering the precessional effects of the spins over this length scale. For this reason, here we consider the full spatial variation of spin accumulation in the nano-pillar, which is calculated at each iteration of the solving protocol by coupling the LLG equation (used for the determinations of the dynamics of the system) to diffusive transport equations. In this picture, the spin accumulation $\vects$ induces an additional contribution to the effective magnetic field acting on the local magnetisation.

The system of equations that have to be solved to determine the spin accumulation $\vects$ is the following \cite{STU2016}:

\begin{eqnarray}
\Div~\vectj=-\Div~C_{0}\;\Grad~\mathrm{\mathbf{V}}=0\;,
\label{eq:divJ}
\end{eqnarray}
\olivierAdd{
\begin{eqnarray}
\frac{ \partial Q_{\alpha k}}{\partial x_k} =  -\frac{s_\alpha}{\tausf}  - \frac{(\vects \times \vectm)_\alpha}{\tausd}.
\label{eq:divQ}
\end{eqnarray}
}
\olivierAdd{In \eqnref{eq:divQ} $\alpha$ index refers to the spin space while $k$ index refers to the real space and $k,\alpha\in\{ x,y,z\}^2$.}
\eqnref{eq:divJ} expresses the total current $\vectj$ resulting from the local electric field $\mathrm{\mathbf{E}}=-\Grad\mathrm{\mathbf{V}}$, $C_{0}$ being the electric conductivity. 
In our convention, positive current $\vectj$ means positive charge flowing from the reference layer to the free layer, which means, electrons flowing from the free layer through the tunnel barrier, and vice versa. Additionally, spin has the meaning of magnetic momentum, \ie, is parallel to the corresponding layer magnetisation. \eqnref{eq:divQ} is a conservation equation, expressing the grounds for the variation of local spin accumulation, resulting either from the spatial variation of the spin-polarized current $\vectQ_{i}$ on the left-hand side, or spin flip and spin precession, on the right-hand side. In this context, $i$ refers to the spatial direction ($i \in \{ x,y,z\}$), $\tausf$ is the spin-flip characteristic time, and $\tausd$ is the $\mathrm{sd}$ coupling characteristic time. 
\olivierAdd{The spin-polarized current is defined as:
\begin{eqnarray}
Q_{\alpha k} = -\frac{\mu_B \; P \; C_0}{|e|} \; u_\alpha E_k -D_0 \; \frac{\partial s_{\alpha}}{\partial x_k}.
\label{eq:spin_pol}
\end{eqnarray}
}
In our simulations we simplify the picture by considering that $P$ has the same value as $\beta$, a parameter used to express the polarisation of conduction electrons injected in the ferromagnet \cite{MOO1999, TAO2024}, and known as the polarisability rate.

When describing spin accumulation we have to consider the surroundings and boundary conditions for the magnetic elements. On the side of the tunnel junction we set the boundary condition of a uniform areal density of current. On the surface between the free layer and the electrode, we describe the spatial variation of spin accumulation in the non-magnetic metal contact, with continuity of both the current density and chemical potential at the interface \cite{VAL1993}. In this framework, to hold equations (\ref{eq:divJ}) and (\ref{eq:divQ}) requires no spin accumulation discontinuity at the interface and an equal normal component of the spin polarized current (the needed stimulus to trigger the magnetisation reversal). The length of the non-magnetic electrode is set to \qty{60}{\nano\meter}, and a uniform voltage is applied as boundary condition of the electrode.



We will see in \secref{sec:3level1} that considering spin accumulation instead of interfacial STT indeed leads to drastic quantitative difference in the case of a 3D nanomagnetic object. 

\subsection{\label{sec:2level2.4}Material parameters}

The magnetic material of the free layer is FeCoB and that of the metal is Cu. \tabref{tab:FeCoB} lists the parameters used in the simulations.

\begin{table}[ht]
\caption{\label{tab:FeCoB} Material parameters used in the simulations. If not stated otherwise the parameters apply both to the magnetic system (MS) and the non-magnetic electrode (NM).}
\begin{ruledtabular}
\begin{tabular}{lcr}
\textrm{Parameter}&
\textrm{Symbol}&
\textrm{Value}\\
\colrule
Saturation magnetisation of MS & $\Ms$ & \qty{1}{\mega\ampere\per\meter} \\
Exchange constant & $A_{\mathrm{ex}}$ & \qty{1.5E-11}{\joule\per\meter}\\
Uniaxial anisotropy & $\Ku$ & \qty{0.96E6}{\joule\per\meter\squared}\\
sf diffusive length & $\lsf$ & \qty{5}{\nano\meter}\\
sd exchange length in MS & $\lsdMS$ & \qty{1}{\nano\meter}\\
sd exchange length in NM & $\lsdNM$ & \qty{0}{\nano\meter}\\
Electric conductivity & $C_{0}$ & \qty{4.0E6}{\per\ohm\per\meter}\\
Number of atoms & $N_{0}$ & \qty{1.344E47}{\per\joule\per\meter\cubed}\\
Polarisability rate & $\beta$ & $0.57$\\
Polarisability ratio & $P$ & $0.57$\\
\end{tabular}
\end{ruledtabular}
\end{table}

\section{\label{sec:3level1}Results}

In this section, we highlight simulation and physical features specifically related to the consideration of a 3D nanosystem, which do not show up when considering the thin free layer of a standard MRAM.

\subsection{\label{sec:3level2.1}Meshing}
A usual concern in micromagnetics is the choice of the meshing size. To investigate this aspect in the present situation, we consider the demagnetising factors $N_{\mathrm{i}}$ ($i \in \{x,y,z\}$). These are important micromagnetic parameters as they determine the internal field for macrospins, and thus their precession frequency. Besides, the stability parameter $\Delta$ of a PSA-MTJ is directly related to the energy barrier of the system $\Eb$, itself linked to $N_{i}$.
Starting from the values of the simulated demagnetising energies $\Ed$ for a uniformly-magnetised state, the demagnetising factors $N_{i}$ are computed as:

\begin{eqnarray}
N_{i}=\frac{2 {\Ed}_{\mathrm{,}i}}{\mu_{0}\Ms^{2}V}~\text{where}~i \in \{x,y,z\}
\label{eq:dem}.
\end{eqnarray}
Since the sum of the three demagnetising factors along the main directions is equal to one ($N_{x}+N_{y}+N_{z}=1$) and given that for a cylinder $N_x = N_y$, we may restrict the study of the numerical evaluation of $\Ed$ in one direction. In principle, the results should be the same no matter the considered direction. Nevertheless we computed  $(N_{x}-N_{z})$ twice, starting either from ${\Ed}_{,x}$ or ${\Ed}_{,z}$. The results are reported in \tabref{tab:dem} for different vertical aspect ratios $\mathrm{AR}=t/D$ of the magnetic cylinder, with $t$ the thickness of the free layer and $D$ its diameter. \olivierAdd{Furthermore, in \tabref{tab:dem} are reported also the values of $(N_{x}-N_{z})$ taking into account the analytical values of the demagnetising coefficients known from literature \cite{ARR1979, SAT1989}.}

\def\epsilonxz{\epsilon_{xz}}

\begin{table}[ht]
\caption{\label{tab:dem} Demagnetising factors of cylinders of different aspect ratio calculated from the simulated value of either ${\Ed}_{,x}$ or ${\Ed}_{,z}$ for uniform magnetisation, with subsequent relative error $\epsilonxz$ in the $(N_{x}-N_{z})$ difference value. The average size of the edge of a tetrahedron constituting the mesh is of \qty{4}{\nano\meter} for all five cases. \olivierAdd{Additionally, the analytical value for the demagnetising factors difference is reported \cite{SAT1989}.}}
\begin{ruledtabular}
\begin{tabular}{l | c c c | r}
\textrm{AR}& &
\textrm{$\mid N_{x}-N_{z}\mid$}& &
\textrm{$\epsilonxz$}\\
  & \olivierAdd{Analytic} & ${\Ed}_{,x}$ & ${\Ed}_{,z}$ & \qty{11}{\%}\\
\colrule
$0.2$ & \olivierAdd{$0.52$} & $0.55$ & $0.49$ & \qty{11}{\%}\\
$1$ & \olivierAdd{$0.032$} & $0.02$ & $0.065$ & \qty{70}{\%}\\
$1.5$ & \olivierAdd{$0.158$} & $0.14$ & $0.161$ & \qty{15}{\%}\\
$2$ & \olivierAdd{$0.227$} & $0.2$ & $0.245$ & \qty{18}{\%}\\
$3$ & \olivierAdd{$0.308$} & $0.296$ & $0.308$ & \qty{4}{\%}\\
\end{tabular}
\end{ruledtabular}
\end{table}

The relative error in \tabref{tab:dem} is computed as:
\begin{eqnarray}
\epsilonxz=\frac{(N_{x}-N_{z})\mid_{{\Ed}_{,x}}-(N_{x}-N_{z})\mid_{{\Ed}_{,z}}}{(N_{x}-N_{z})\mid_{{\Ed}_{,x}}}
\label{eq:rel_err}
\end{eqnarray}
The relative error for the cylinder with AR~$1$ is striking. It reaches about $70\%$, while it drops sharply for larger AR. This can be understood as $1$ is an AR for which the magnetisation orientation in the system is close to the transition from out-of-plane to in-plane for lower values, so that small errors translate into large differences. The error drops to $26\%$ upon halving the average length of the edge of tetrahedron unit cells to \qty{2}{\nano\meter}. This issue of large relative error is specific for a 3D nano-object, as for the flat shape of standard p-MRAM the $x$ and $z$ demagnetising coefficients are very different one from another, so that errors in the difference are low. This shows that one needs to pay extra attention to mesh size refinement in 3D.

\subsection{\label{sec:3level2.2}Flower state}

\begin{figure}[ht]
	\begin{center}
	\includegraphics[scale=0.45]{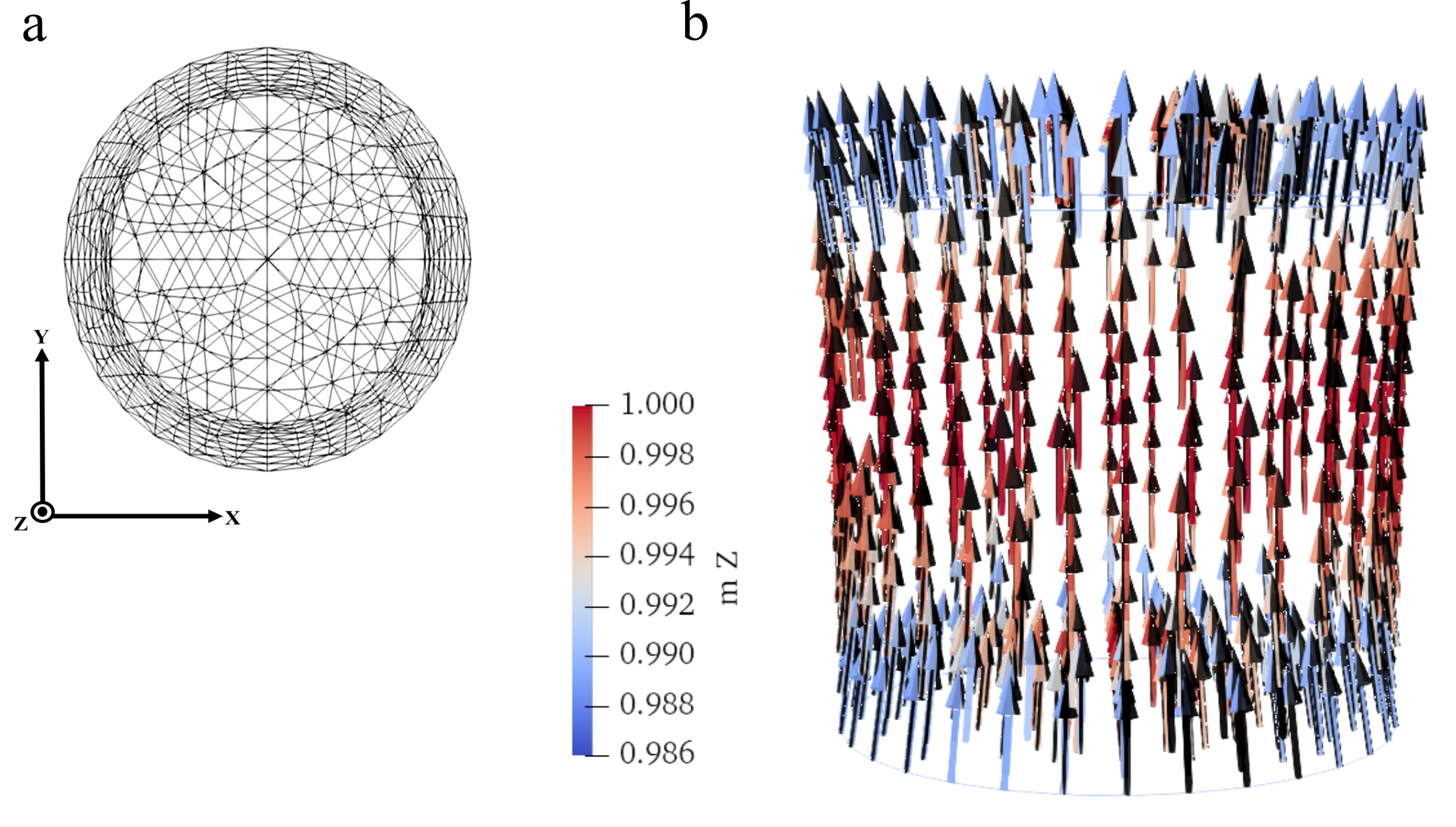}
		\caption{(a) Representation of the FEM mesh used to discretize a pillar with \qty{20}{\nano\meter} height and \qty{20}{\nano\meter} diameter. (b)~Relaxed micromagnetic state. The arrows represent the direction of the local magnetic moments. In order to highlight the flower-state a color scale indicates the value of the $z$ component of the unit magnetisation.} \label{fig:flower-state}
	\end{center}
\end{figure}

\figref{fig:flower-state} shows the micromagnetic state in a pillar with vertical aspect ratio$~1$, after relaxation taking from uniform magnetisation as the starting point. There are deviations of magnetisation from the uniform state at the edges of the structure, for both the upper and lower surfaces, which is called a flower state\cite{SCH1988}. This situation results from the divergence of dipolar fields at edges and corners\cite{RAV1998}, whose cost in terms of dipolar energy is decreased by rotating magnetisation towards the diagonal to the edge, at the expense of exchange energy\cite{SLA2010}. In our case the angular deviation from the $\mathrm{z}$-axis at the very edge is about \qty{9}{\degree}, while at the center of both surfaces magnetisation is aligned with the $\mathrm{z}$-axis. The occurrence of the flower state is specific to a 3D system, with a size larger than the dipolar exchange length. In the thin free layer of an MRAM the dipolar effects from top and bottom surfaces overlap and cancel each other at the length scale of the layer thickness, leaving it in a uniform state. We will see in \secref{sec:3level2.3} that this has a direct and specific effect on current-driven magnetisation dynamics.

\olivierAdd{The question if the flower state survives temperature effects and interfacial effects (such as the interfacial  perpendicular  magnetic  anisotropy or the  Dzyaloshinskii-Moriya interaction) could spontaneously rise. We discuss thoroughly those points in the appendix \secref{sec:5level1}. However we may already anticipate that both aspects enhance or reverse the flower state into a leaf state, with an expected impact on the incubation time of the magnetisation reversal (deeply analyzed in \secref{sec:3level2.3}).}

\subsection{\label{sec:spinAccumulation}Spin accumulation}

\begin{figure}[ht]
	\begin{center}
		\includegraphics[scale=0.35]{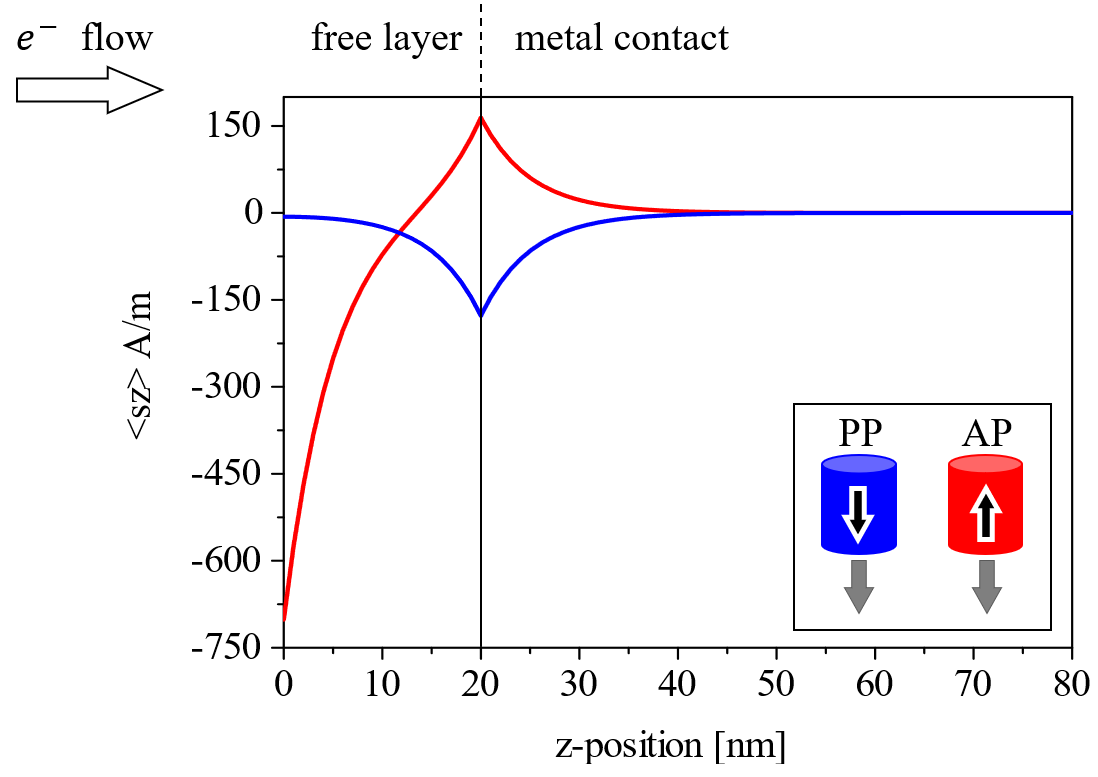} 
		\caption{Spin accumulation in the magnetic layer and metal contact for a free layer of \qty{20}{\nano\meter} both thickness and diameter. The current density injected through the tunnel barrier is $j=$\qty{-5E12}{\ampere\per\meter\squared}. The vertical axis represents the cross-sectional-average of the $z$ component of spin accumulation. The situations for the parallel state of uniformly-magnetised polarizer and free layer are depicted in blue and red, respectively. The inset illustrates the geometry, of the parallel (PP) and anti-parallel (AP) configurations with positive charge current, gray arrows representing the magnetisation direction in the reference layer and the black arrows the magnetisation direction in the free layer.} \label{fig:boundary_conditions}
	\end{center}
\end{figure}

\figref{fig:boundary_conditions} illustrates the simulated spin accumulation $\vects$ in the magnetic and the metal parts for a structure with aspect ratio $\mathrm{AR}=1$. The blue and red curves represent the spin accumulation for the parallel and anti-parallel states, respectively. As mentioned in the methods part, the so-called spins are parallel to the associated magnetic moment, and electrons flow in the direction opposite to the charge current. \olivierAdd{\figref{fig:boundary_conditions} is drawn for negative current, \ie, electrons are injected from the polarized layer into the free layer.} Several aspects are worth mentioning. First, spin accumulation occurs not only at the tunnel barrier, but also at the interface with the normal metal. In a metal/ferromagnet/metal stack this would induce an antidamping-like torque at the interface where the electrons enter the ferromagnet due to the accumulation of spins antiparallel to magnetisation, and a damping-like torque at the interface where the electrons leave the ferromagnet, due to the accumulation of spins parallel to magnetisation. This picture is consistent with the generalized Landau-Lifshitz-Gilbert-Slonczewski equation\cite{THI2005}, if the gradient of magnetisation close to the interfaces is considered: the flow of electrons promote a larger angular deviation at the interface through which they enter the mostly uniformly-magnetized material, and a lower angle at the interface through which they exit the material. Second, STT is not localized at interfaces, but decays progressively away from these at the length scale of the spin diffusion length, such as described by the Valet-Fert model for giant magneto-resistance in the current-perpendicular-to-plane geometry\cite{VAL1993}. This length scale is a few nanometers, comparable to magnetic length scales such as the dipolar and exchange length, and therefore liable to efficiency couple to non-uniform magnetisation distributions such as the flower state described previously. Third and as a consequence of the previous statement, the height of the pillar has a direct impact on the value of spin accumulation, depending whether it is smaller or larger than the spin diffusion length. The former case is that of a standard MRAMs with a thin free layer, for which the Valet-Fert-like accumulations with opposite signs overlap and largely cancel each other, leaving the polarisation of the injected current as the main driving force for spin accumulation, independent of the magnetisation direction in the free layer. For the latter case, which is the situation of study here, the magnitude of spin accumulation at the tunnel barrier depends strongly on the direction of magnetisation of the free layer. 

\begin{figure}[h]
	\begin{center}
		\includegraphics[scale=0.15]{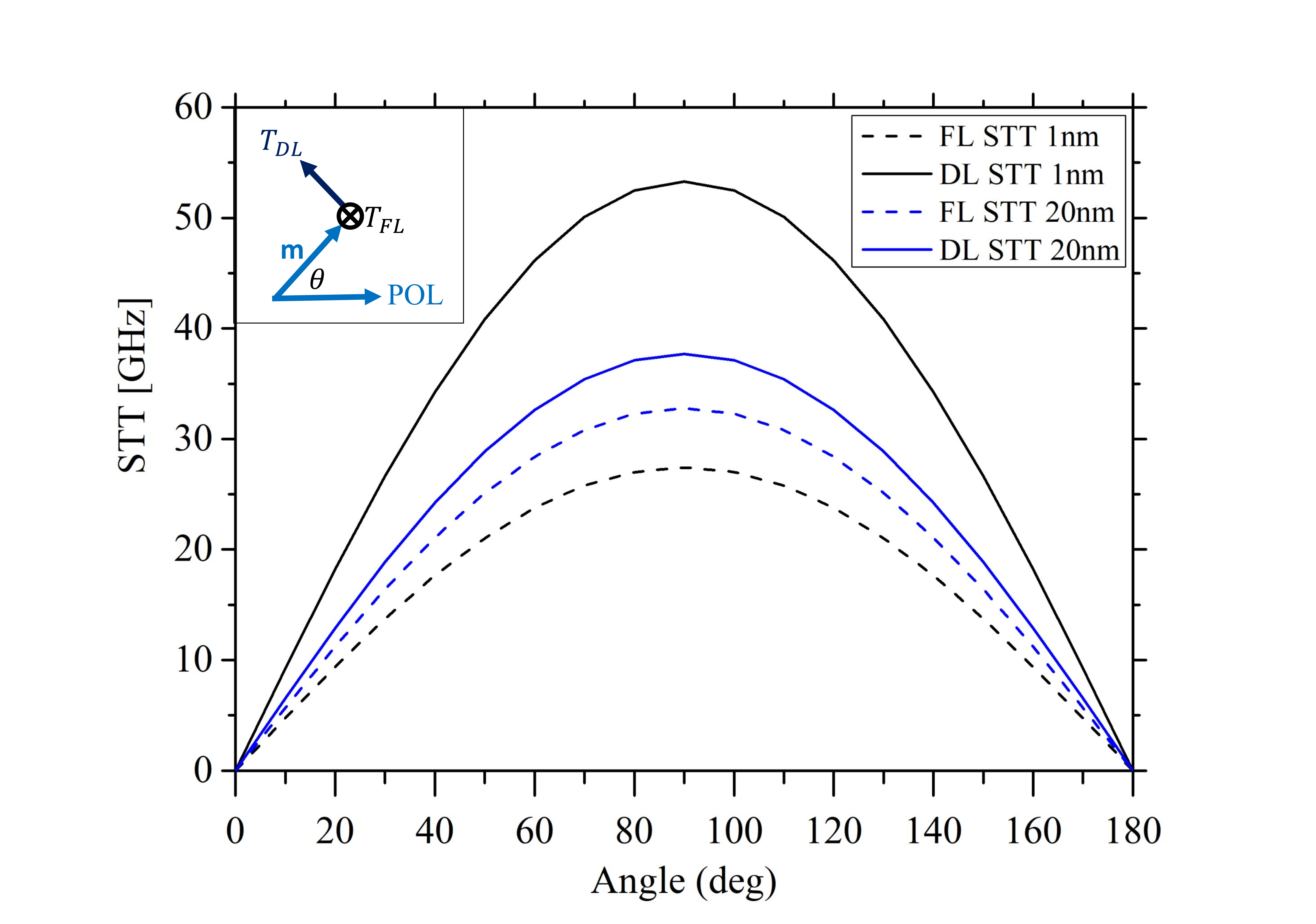} 
		\caption{STT components versus the angle of magnetisation in the free layer, with respect to the $z$-axis. Black stands for a \qty{1}{\nano\meter}-thick system, while in blue stands for a \qty{20}{\nano\meter}-thick system. FL torques are shown with full lines, DL torques with dotted lines. } \label{fig:STTcomponents}
	\end{center}
\end{figure}

Another difference between the model of effective STT and the one of spin accumulation is that the ratio of damping-like and field-like torques is a direct output of the latter. To disentangle both we compute the amplitude of STT versus the misalignment angle of magnetisation with the $z$-axis, while keeping the polarizer fixed\bracketfigref{fig:STTcomponents}. The results for a 2D-like system (cylinder of \qty{1}{\nano\meter} $\times$ \qty{20}{\nano\meter}) and a 3D-like one (cylinder of \qty{20}{\nano\meter} $\times$ \qty{20}{\nano\meter}) are numerically computed, taking into account the value of the torque in reduced units (see equation \ref{eq:T_STT}).
\bracketfigref{fig:STTcomponents} shows that while in the case of the 2D system the DL component dominates over the FL, as has been pointed out already \cite{TIM2015}, the two torques have a similar magnitude for the 3D system. How to link quantitatively the interfacial and bulk descriptions of STT requires further investigations, which lies beyond the scope of the present work. Nevertheless, the effects mentioned above already show qualitatively what is specific in a 3D nanomagnetic system, and expected to give rise to peculiar magnetisation dynamics, which we explore in the next paragraph.

\subsection{\label{sec:3level2.3}Current-driven magnetisation reversal}

If not stated otherwise, simulations are conducted for a cylinder having a diameter of \qty{20}{\nano\meter} and a thickness of \qty{20}{\nano\meter}, and a Gilbert damping factor $\alpha=0.02$. The starting point of the simulation is uniform magnetisation with $1^{\circ}$ misalignment with respect to $\mathrm{z}$-axis, and abrupt flow of electric current from the start. Some of the values for the density of injected current are unrealistically-high for an experimental current through a tunnel barrier. However, covering a large range of stimulus allows us to highlight the various possible physics of 3D systems and their subsequent potentialities.\\

\begin{figure}[ht]
	\begin{center}
		\includegraphics[scale=0.25]{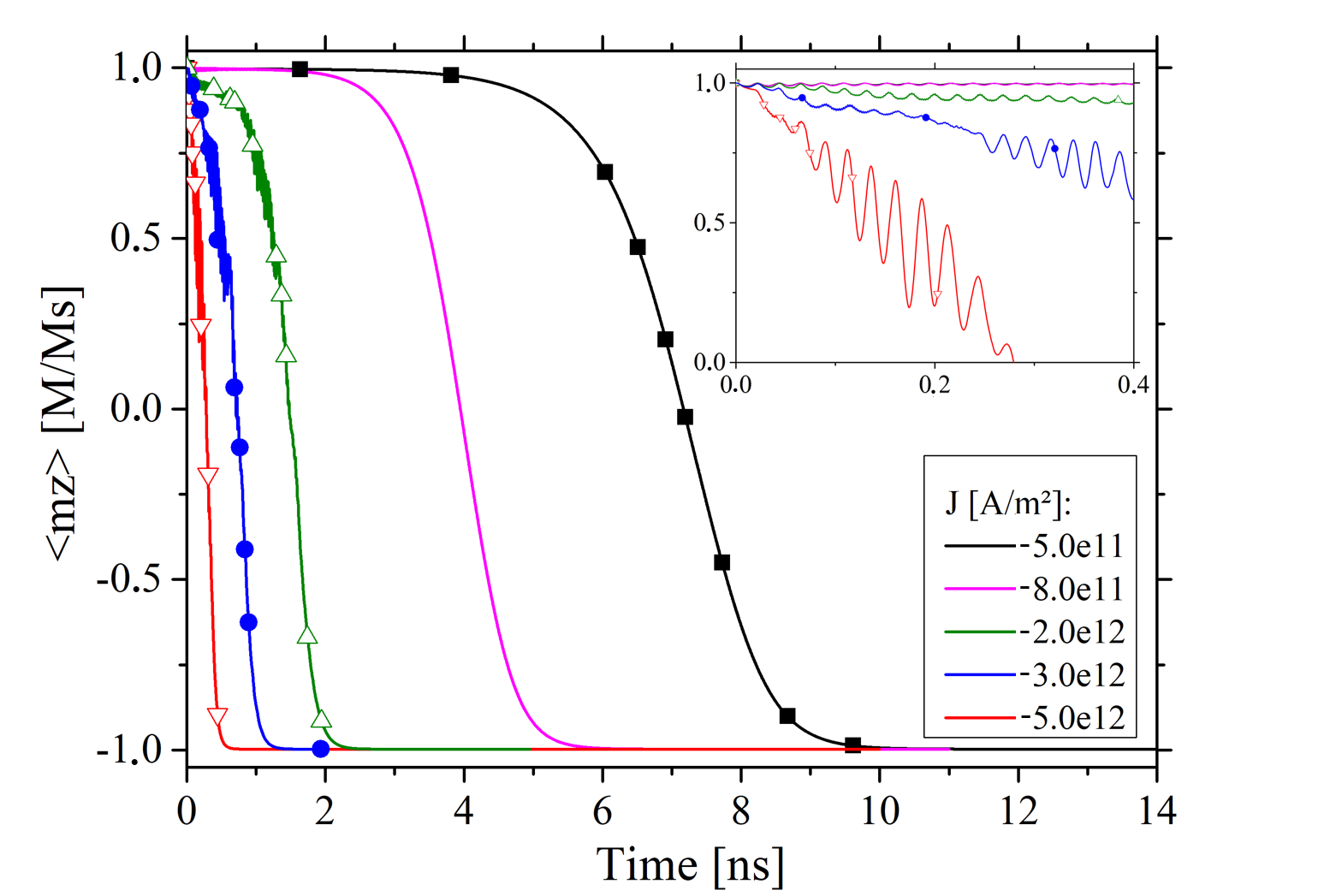} 
		\caption{$z$ component of the unit magnetisation vs time for different values of injected electric current densities. The symbols refer to the time positions at which the snapshots reported in \figref{fig:reversal-mode} were taken. The inset is a close-up view of the magnetisation evolution during the first \qty{0.4}{\nano\second}, highlighting oscillations.} \label{fig:reversal-speed}
	\end{center}
\end{figure}

\def\tauUp{\tau_\mathrm{i}}%
\def\tauDown{\tau_{95}}%
\def\tauSwitch{\tau_\mathrm{s}}%

\figref{fig:reversal-speed} shows the evolution of the $z$ component of the average unit magnetisation $m_{z}$ versus time, for five different values of injected current densities. $m_{z}=\pm1$ means uniform magnetisation aligned with the $z$-axis upwards and downwards, respectively. Let us define two quantities to analyse the time scales of magnetisation reversal. A first time scale appearing on \figref{fig:reversal-speed} is the time needed before a sizable change is noticeable. We define it arbitrarily when $m_{z}$ reaches $0.9$, which we call $\tauUp$, It is sometimes called the incubation time, which for a macrospin reflects the gradual rise of the precession angle under the influence of the antidamping torque. It is usually strongly dependent on initial misalignement of the various magnetisation vectors, and on thermal and numerical noise.  A second time scale is the time needed for the magnetic moment to go from mostly up to mostly down. We define it the time for $m_{z}$ to go from $+0.9$ to $-0.9$, which we call $\tauSwitch$. It is sometimes called the switching time.

\begin{figure}[h]
	\begin{center}
		\includegraphics[scale=0.3]{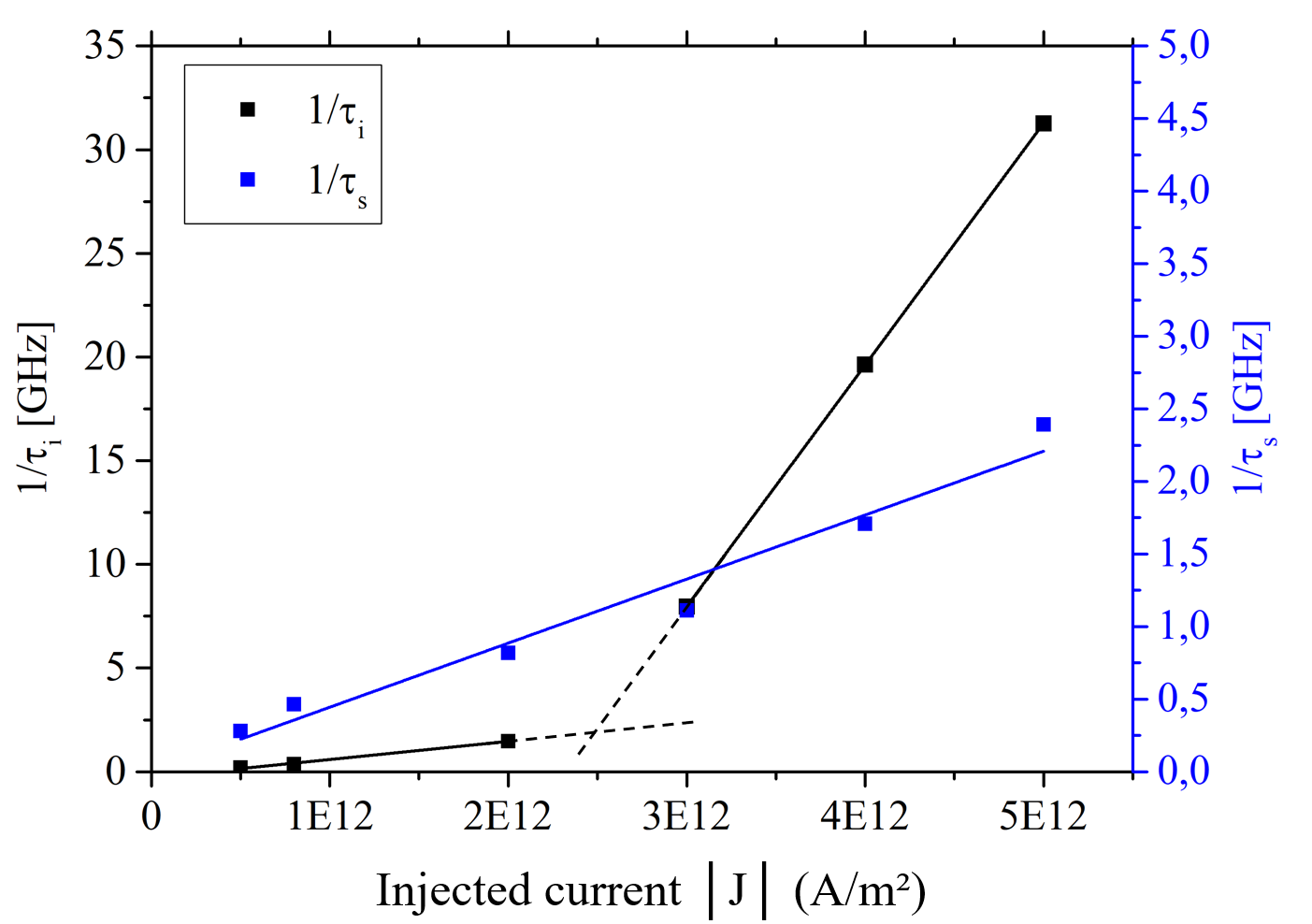}
		\caption{Evolution of the characteristics inverse times $1/\tauUp$ (black curve) and $1/\tauSwitch$ (blue curve) with the absolute value of the injected STT current. Two different scales are applied for the two curves.} \label{fig:1/tau-vs-j}
	\end{center}
\end{figure}

\figref{fig:1/tau-vs-j} shows that up to a current density $\mid j \mid$ of circa \qty{1E12}{\ampere\per\meter\squared}, $\tauUp$ vary inversely linearly with the injected current. This behavior is expected\cite{SUN2000} and observed\cite{BED2010} for the spin-transfer switching of a macrospin when the current exceeds the threshold for switching, simply reflecting the rate of transfer of angular momentum from spin accumulation to magnetisation\cite{SLO1996}. The low slope is predicted to reflect in a logarithmic manner the initial misalignment angle~$\theta_0$. $\tauSwitch$ also varies inversely linearly with injected current however with a larger slope, reflecting the larger reference angle, with cosine equaling $0.9$.

However, two features suggest that the the magnetisation dynamics is not fully grasped by the macrospin picture. First, the inset of \figref{fig:reversal-speed} shows that from the very first stages $m_{z}$ oscillates during the incubation stage. The magnitude of oscillations increases sharply and non-linearly with the current density, and the frequency is more than one order of magnitude larger than the macrospin gyroscopic precession \figref{fig:frequencies}. 
Second, a cross-over occurs at large current for the variation of inverse $\tauUp$ with applied current. In this regime, $m_{z}$ departs linearly from saturation from the start of simulation, which again does not fit the standard picture of incubation of a macrospin. Note that as \figref{fig:reversal-speed} does not display a plateau at mid-reversal one can probably rule out a mechanism of nucleation-propagation of domain wall\cite{CAC2021, TAK2015}. In the following paragraph we examine the magnetisation distributions during reversal to understand the microscopic reasons for these features.

\begin{figure}[h]
	\begin{center}
		\includegraphics[scale=0.42]{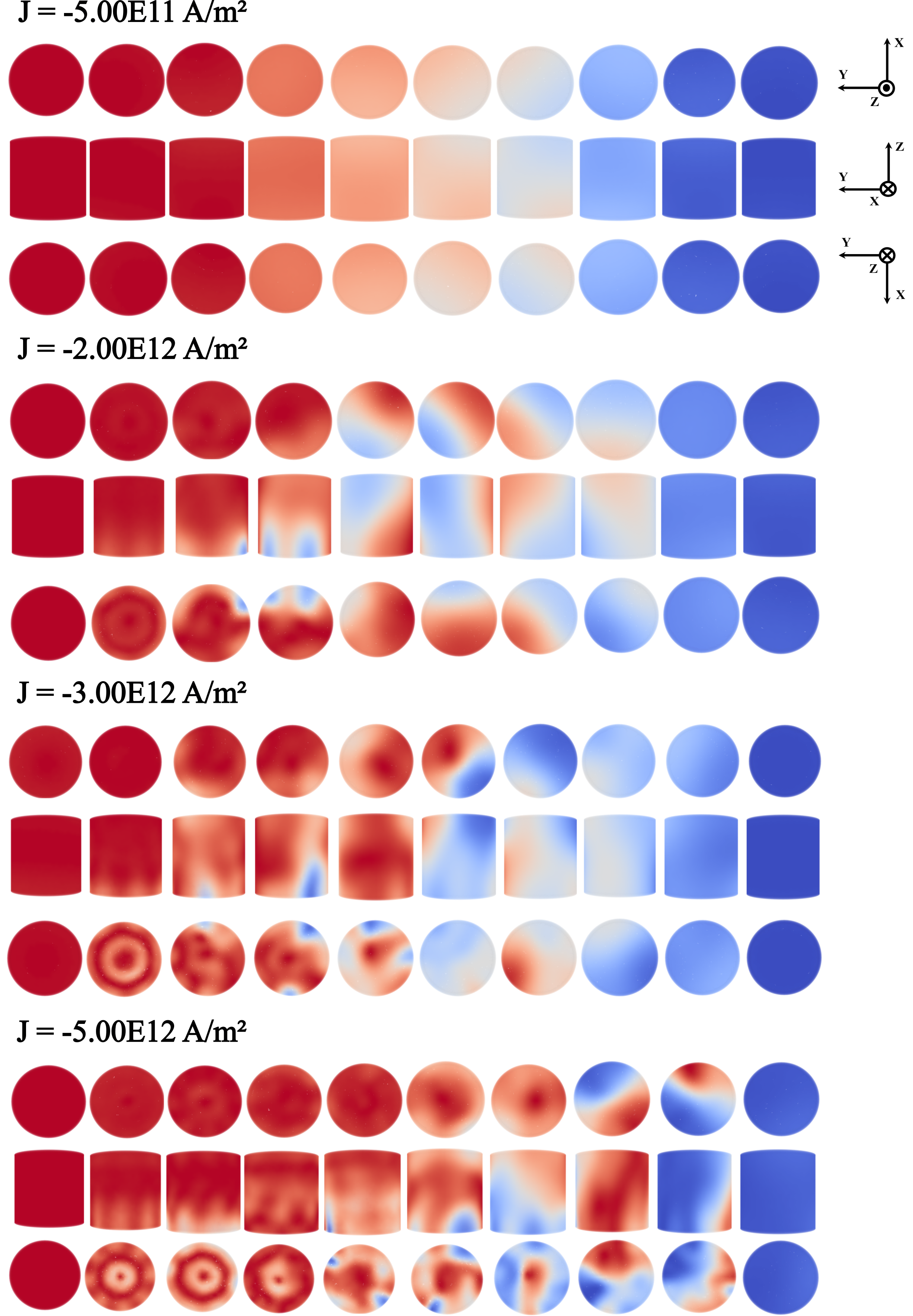} 
		\caption{Top, side and bottom views of magnetisation at the surface of a cylindrical volume of \qty{20}{\nano\meter} thickness and \qty{20}{\nano\meter} diameter over the current-driven switching process. Red corresponds to upward pointing magnetisation while blue corresponds to downward pointing magnetisation. From top to bottom are represented the results for increasing injected current densities. The different snapshots are taken for increasing time steps, starting from \qty{0}{\nano\second}. The time steps are not the same in the four cases but are spread over the switching time required by the specific configuration. The time steps at which the snapshots are taken are highlighted with symbols on the curves of \figref{fig:reversal-speed}.} \label{fig:reversal-mode}
	\end{center}
\end{figure}

\figref{fig:reversal-mode} illustrates the magnetisation distribution at the surface of the magnetic cylinder during the reversal process, for different values of injected current. This series highlights a cross-over for nearly coherent reversal at low current density to incoherent at larger current density, with a threshold value comparable to the cross-over of $1/\tauUp$ in \figref{fig:1/tau-vs-j}. Deviations from uniform magnetisation appear first at the edges of the interface with the tunnel junction, an effect most clearly visible for current density \qty{-3E12}{\ampere\per\meter\squared} and higher. The linear decrease of $m_z$ versus time observed for the highest current densities in \figref{fig:reversal-speed} is explained by this non-uniform excitation mode, as the angle of precession can reach up to several tens of degrees and involve a macroscopic fraction of the volume of the cylinder. The excitation of these modes can be understood as spin accumulation is largest at this interface, and its effect is most efficient at the edges of the pillar due to the flower state. The efficient pumping of energy at these locations is confirmed by movies of the full magnetisation evolution, revealing the rise of precession of magnetisation turning the micromagnetic state periodically in flower-like, and leaf-like configurations with intermediate curling configurations with opposite chiralities~(See animations in supplementary material\cite{SUPP2025}), with a frequency linked and thus explaining the oscillations of $m_z$~(inset of \figref{fig:reversal-speed}). The frequency of precession $\fedge$ is in the range \qtyrange{40}{45}{\giga\hertz} (\figref{fig:frequencies}). This is fully inconsistent with the Kittel FMR formula of a macrospin in its own demagnetising field, related to $N_{x}-N_{z}$\cite{KIT1948}:
\begin{eqnarray}
f=\frac{\gamma_{0}}{2\pi}\sqrt{(N_{x}-N_{z})\Ms}\;,
\label{eq:kittel}
\end{eqnarray}
which amounts to \qty{1.23}{\giga\hertz}. This discrepancy and the localisation of the non-uniform mode at the edges of the cylinder makes clear that the local effective field involved is not the demagnetising field, but a combination of dipolar and exchange field involved in the flower state. The oscillation frequencies convert to $\approx$\qty{1.3}{\mega\ampere\per\meter} in terms of effective induction field. Given the fact that the radius of the cylinder is close to four times the dipolar exchange length, this is exactly the range of frequency expected from the interplay of magnetostatics and exchange at this scale. Examining the evolution over time for a given current density is also interesting. It is clear that a breaking of symmetry is needed to reverse a macrospin, and this is what happens for the low current density. A breaking of symmetry is also required even in the case when an edge mode is the way that the system gets efficiently excited, for the large current densities. Indeed, magnetisation reversal via a perfectly-symmetry curling state requires the nucleation of a Bloch point on the axis of the cylinder\cite{ARR1979b}, which is an energetically costly process\cite{THI2003}. Examining the edge modes over time, we notice that the breaking of symmetry takes the form of the mode changing progressively from a high azimuthal one to a low-azimuthal mode. 

The change of azimuthal mode is associated with a transient state with low amplitude of oscillation, before oscillations are recovered however with a slightly larger frequency increase, see for instance \figref{fig:reversal-speed} for current density \qty{-3E12}{\ampere\per\squared\meter}.

Let us finally discuss the dependence of $\fedge$ with applied current, which decreases by about \qty{6}{\giga\hertz} from low current to \qty{-5E12}{\ampere\per\meter\squared}, so, amounting to \qty{1.2}{\giga\hertz\per\ampere\meter\squared}. Two phenomena may explain this dependence. The first possibility is the non-linearity of precession\cite{RUI2017}, due to the angle reaching several tens of degrees shortly before reversal for the largest current density. However, there is no clear gradual change of frequency during the increase of amplitude of oscillation for a given current density. The second possibility is the effective field arising from the field-like component of STT, the decrease being consistent with the expected antidamping effect. We may link this to the macrospin oscillations observed in the case of a damping torque. For a cylinder displaying \qty{20}{\nano\meter} thickness and returning to equilibrium, the precession frequency changes from \qty{2.13}{\giga\hertz} at zero injected current to \qty{2.35}{\giga\hertz} for an injected current of \qty{-6.79E11}{\ampere\per\meter\squared}, \ie, amounting to $\approx$\qty{0.3}{\giga\hertz\per\ampere\meter\squared}. Both figures are therefore not consistent. In the end, there may be a contribution of both effects, non-linearity and field-like torque. The non-linearity of precession is also expected to induce a non-linear effect on the effective field-like torque, as it is the exchange between spin accumulation and magnetisation that connects the damping-like and field-like terms in our approach (see equation \ref{eq:T_STT}).

\begin{figure}[h]
	\begin{center}
		\includegraphics[scale=0.25]{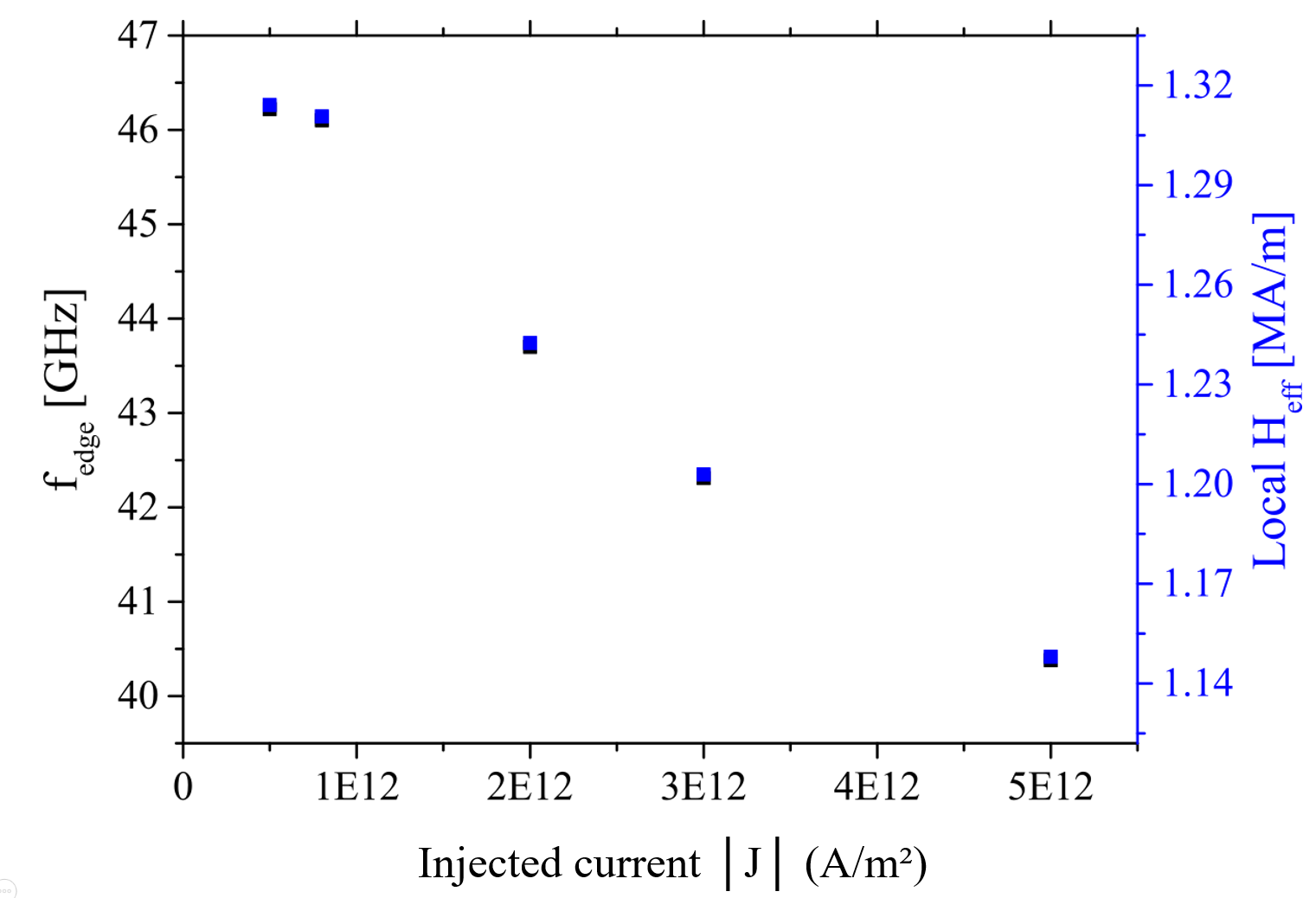} 
		\caption{Frequency of oscillation of $m_z$ derived from the inset of \figref{fig:reversal-speed}, expressed in \si{\giga\hertz} and in the equivalent effective field for gyrotropic motion $H_{\mathrm{eff}}$.} \label{fig:frequencies}
	\end{center}
\end{figure}

To conclude this part, edge modes are efficiently excited in 3D nanosystems, due to the combination of the flower state and larger field-like torque. This leads to pathways for magnetisation reversal which may deviate from the macrospin picture at large driving currents. \\
\section{\label{sec:4level1}Conclusions}
We explored the details of spin-transfer-torque switching mechanism of 3D nanocylinders connected with a magnetic tunnel junction, and highlight the physical phenomena specific to the 3D situation, compared with thin elements such as found in standard MRAM. First, we showed that the explicit consideration of spin accumulation is necessary, being sizable in the vicinity of both interfaces, with the tunnel barrier and the metal lead. Besides, the field-like term is comparable with the damping-like one, contrary to the 2D case. This, combined with the flower state naturally occurring in a 3D element, makes the STT very efficient at the edges of the cylinder and excites edge precession modes with large magnitude and high frequency. These play a key role in the reversal, inducing in particular a quick and linear decrease of longitudinal magnetic moment versus time instead of the long incubation stage in a flat cell, which is liable to accelerate magnetisation reversal. We believe that our results are meaningful when optimizing perpendicular-shape-anisotropy MRAM cells, and beyond, should be applicable to many situations in the emerging field of 3D spintronics.

\begin{acknowledgments}
The authors would like to thank Mairbek Chshiev, Christophe Thirion, Ricardo Sousa and Nuno Caçoilo for fruitful discussions. This work has been partially supported by the ANR project M-bed-RAM ANR-23-CE24-0016.
\end{acknowledgments} 

\section{\label{sec:5level1}Appendix}
\subsection{Thermal effects}
\begin{figure}[ht]
	    \begin{center}
		    \includegraphics[scale=0.15]{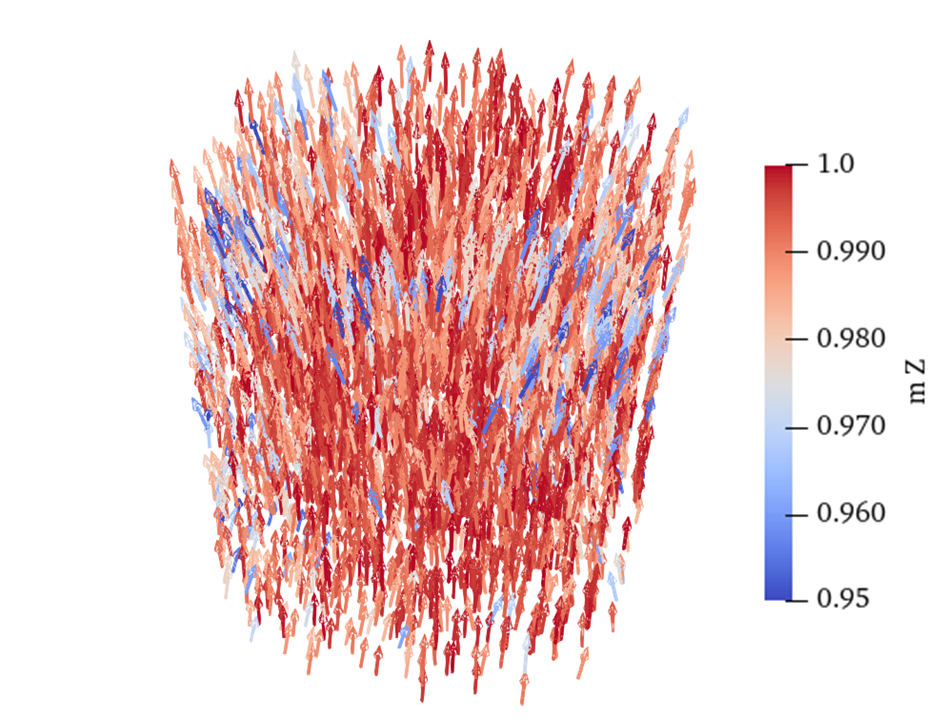} 
		    \caption{Relaxed state of a nanomagnetic cylinder when the environmental temperature is assumed to be \qty{50}{\kelvin}. } \label{fig:T_relax}
	    \end{center}
    \end{figure}

Thermal fluctuations have a significant impact on the physics of magnetisation reversal in nanomagnetic systems. It is common knowledge that their impact increases for low-dimensionality or low-volume systems, decreasing the energy barriers preventing switching. This leads to the effect of superparamagnetism. In MRAM, the decrease of energy barriers shows up for thin devices with diameter typically below the usual ones for MRAM\cite{PER2018, WAT2018}. As expected, energy barriers are boosted by the much increased volume. Thanks to this structural change, PSA-STT-MRAMs exhibit much greater resistance to thermal events. As thoroughly in\cite{CAC2023(Cacoilo-PhD-PSA)}, thermal fluctuations mainly affect the incubation time in the total magnetisation reversal mechanism of PSA-STT-MRAMs, promoting symmetry breaking. However, no unwanted switching at zero applied voltage is observed, thanks to the thermal retention provided by the perpendicular shape anisotropy. The quantitative imaging of magnetisation in the free layer of a PSA MRAM directly confirmed the reduced impact of thermal energy in these.
As far as the flower state is concerned, as it arises in a reduced volume compared to the entire pillar, it is indeed relevant to question the role of temperature. To do so, we performed additional simulations using a finite-difference effective-STT code, which considers thermal noise with the Langevin equation, magnetisation being assumed to be in contact with a heat bath at temperature $T$. We considered an environmental temperature of \qty{50}{\kelvin}, which is expected to correctly describe the experimental situation at \qty{300}{\kelvin}, as the Langevin approach with macrospins overestimates the impact of temperature \cite{EVA2015, WOO2015}. 
The results are shown in Figure~\ref{fig:T_relax}.
    
\begin{figure}[ht]
	    \begin{center}
		    \includegraphics[scale=0.12]{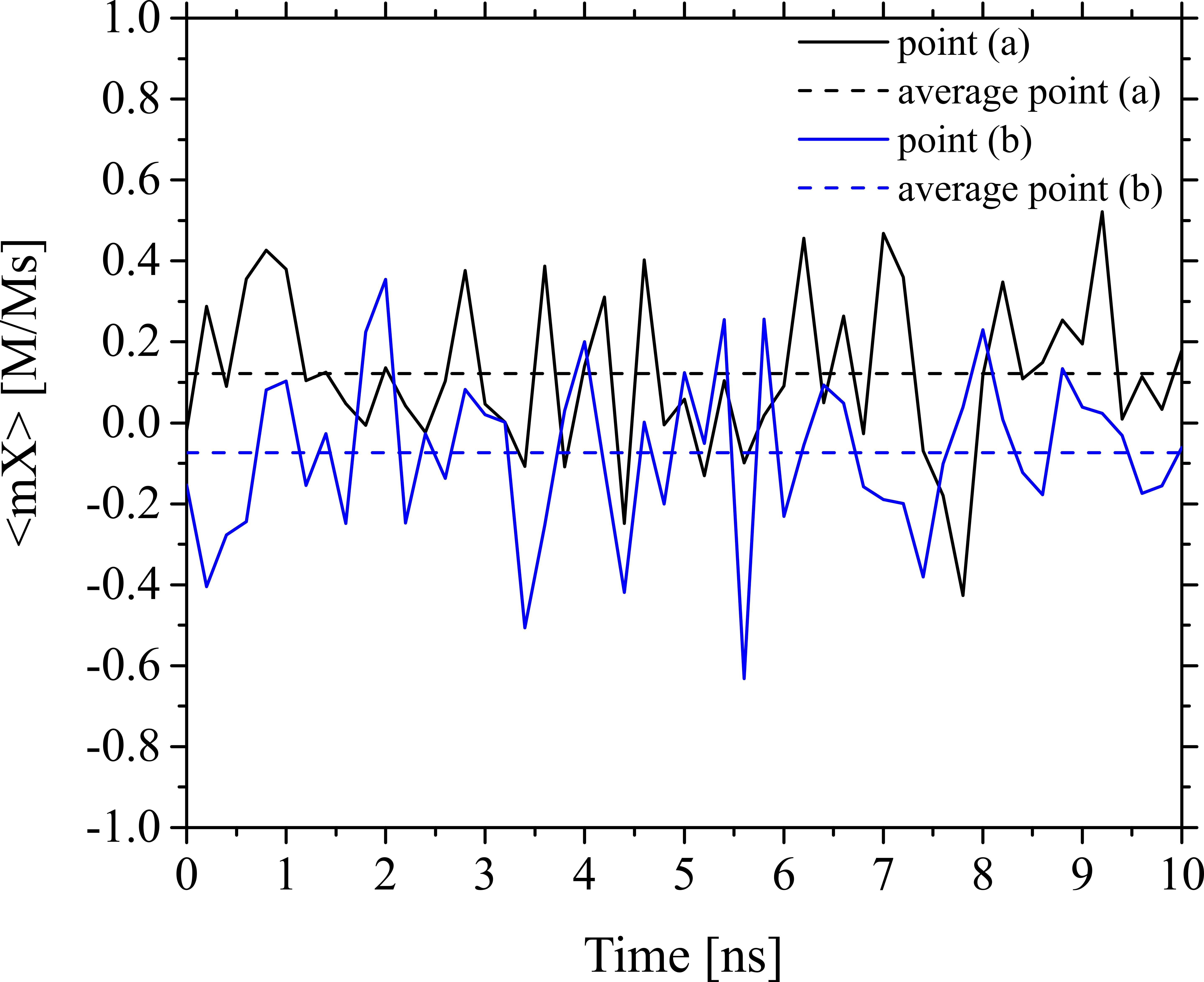} 
		    \caption{Two points along the same diameter on the bottom surface of the investigated cylinder. The average value of the $x$-component of the unitary magnetisation is extracted for both of them on a time interval of \qty{10}{\nano\second}. } \label{fig:mx}
	    \end{center}
    \end{figure}

\figref{fig:T_relax} suggests that the flower state remains despite thermal noise. To confirm this quantitatively, let us examine the average magnetisation at both a central cross-section and at an outer surface cross-section over a time interval of \qty{5}{\nano\second}. We find that the normalized z-component of the magnetisation is $0.993$ at the center plane of the cylinder and $0.985$ at the current-injection surface. In more detail, let us examine the sign of the average $m_x$ component at two points located on opposite sides of a diameter at the outer surface. The extracted values are averaged over a \qty{10}{\nano\second} time interval. For point (b) in\figref{fig:mx} the value is of $-0.07$ (vs $-0.11$ at \qty{0}{\kelvin}) while for point (a) the value is of $0.12$ (vs $0.11$ at \qty{0}{\kelvin}). This demonstrates that the tilt of magnetisation at the edges of the cylinder are preserved on the average, \ie, the flow state survives thermal excitations. Due to the rotational symmetry of the system's physics, it is sufficient to perform this check along a single diameter rather than all possible ones. 
    
\subsection{Interfacial effects}

\begin{figure}[ht]
	    \begin{center}
		    \includegraphics[scale=0.15]{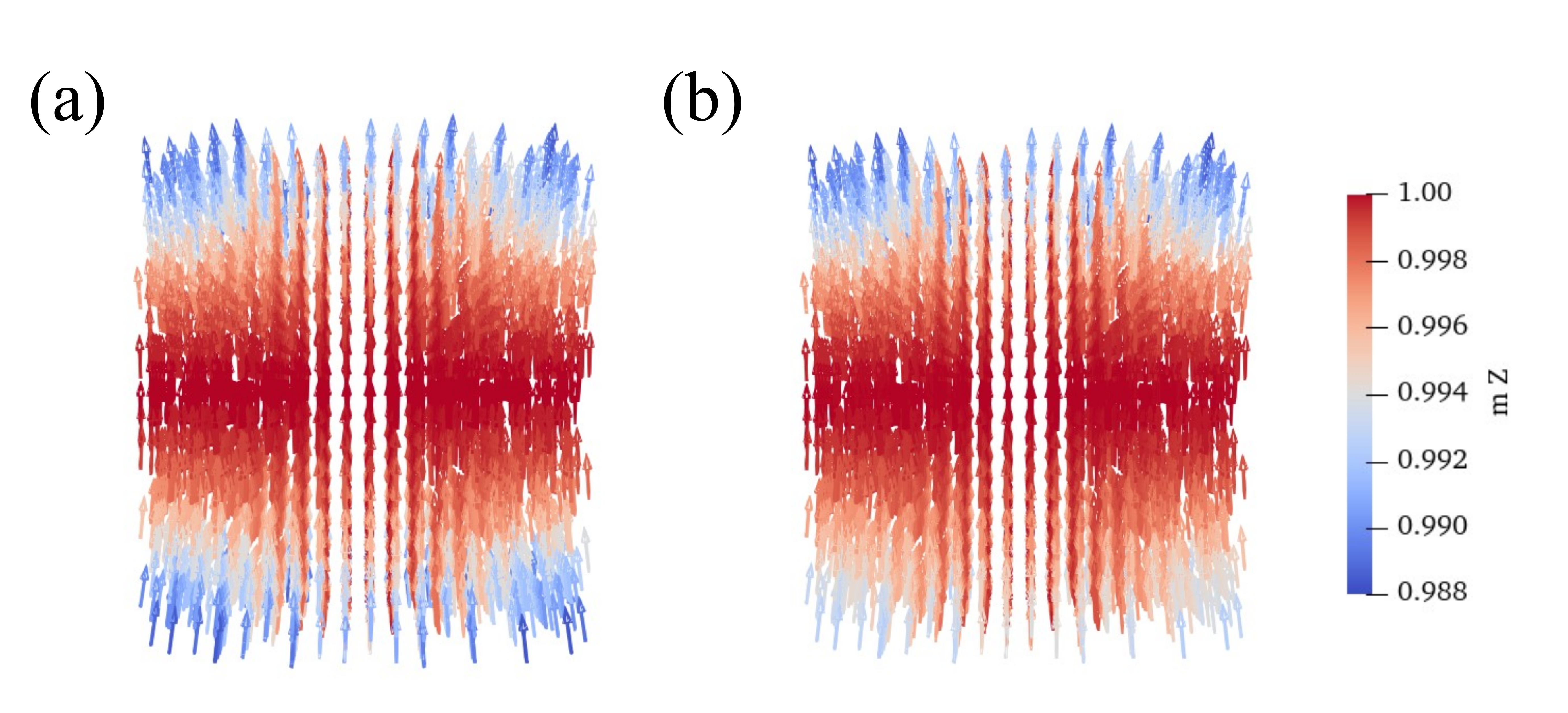} 
		    \caption{Simulation of a nanocylinder without (a) and with (b) iPMA applied to the bottom surface with interfacial anisotropy of $K_\mathrm{s}=$ \qty{1.4}{\milli\joule\per\meter\squared}. } \label{fig:iPMA}
	    \end{center}
    \end{figure}

Various effects may arise at the interfaces between the layers that constitute the structure. Two main types of contributions can be considered: one due to interfacial perpendicular magnetic anisotropy (iPMA) and another due to the Dzyaloshinskii-Moriya interaction (DMI). Both of these effects can influence the flower state observed in the free layer under relaxed conditions. \\
The role of iPMA has already  been considered in nanocylinders, \eg, in Section 2.2 of \cite{CAC2023(Cacoilo-PhD-PSA)}, where iPMA is introduced at the interface between the tunnel barrier and the free layer. 
Using the same finite-difference effective-STT code as for the thermal analysis, let us  introduce in our system an interfacial anisotropy of $K_\mathrm{s}=$ \qty{1.4}{\milli\joule\per\meter\squared}. This value is typical for a FeCoB/MgO interface\cite{PER2018}. As could be expected, iPMA stiffens magnetisation along the perpendicular direction. This decreases the magnitude of the flower state, however there is still a deviation of magnetisation from the $z$~axis, highlighted by a change in the color of the magnetisation vectors at the edge of the structure in \figref{fig:iPMA}.
    
On the other hand, the role of DMI in the flower state of a nanocylinder has not been reported yet, to our knowledge. In \cite{CHE2021}, we find that the largest value reported for the DMI at an MgO/FeCoB interface is \qty{1.13}{\milli\joule\per\meter\squared}. First of all, let us note that the sign of the interaction is positive. This implies that the rotation of the magnetisation is counter-clockwise at the edges of the domain \cite{GUE2025}, \ie, in the same orientation as that of the flower state at the current-injection surface resulting from the dipolar energy contribution. Thus, we would expect that the DMI enhances the flower state at the bottom surface, while disrupts it at the top surface.

\begin{figure}[ht]
	    \begin{center}
		    \includegraphics[scale=0.2]{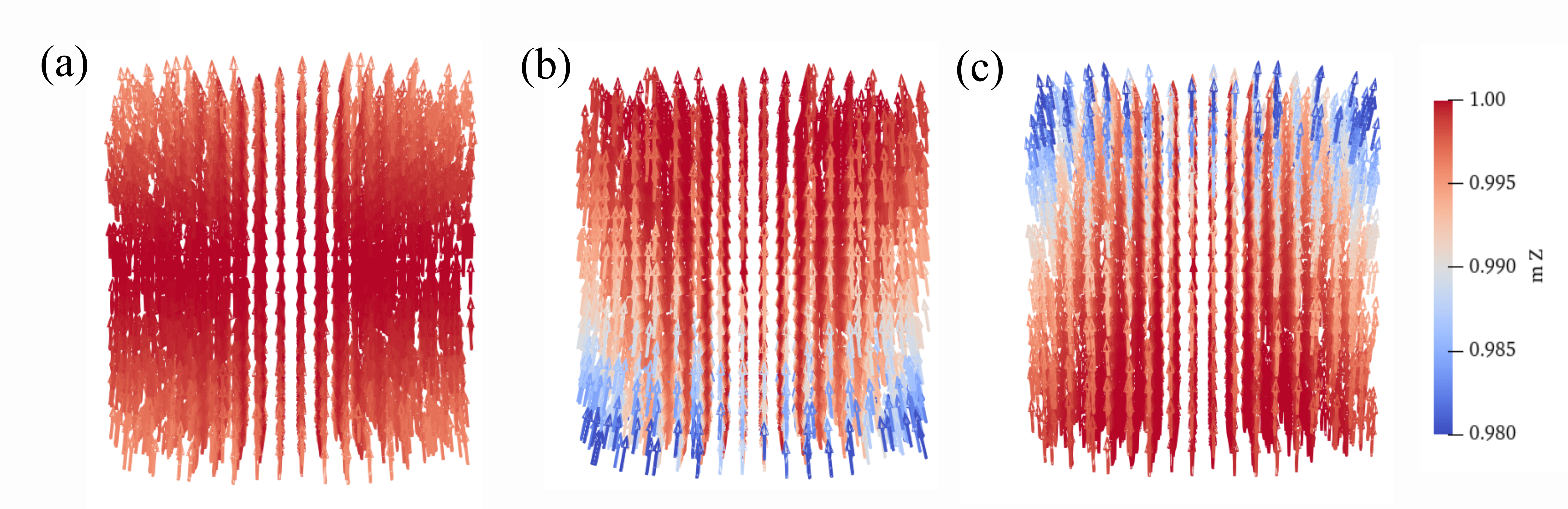} 
		    \caption{Simulation using mumax considering no DMI contribution (a), a positive contribution only at the bottom interface (b) and a negative DMI contribution at bottom interface (c). The DMI constant is taken to be \qty{1.13}{\milli\joule\per\meter\squared} \cite{CHE2021}. } \label{fig:DMI}
	    \end{center}
    \end{figure}

To check this claim, we performed a simulation with mumax, which allows one to account for DMI at the interfacial level. The sign of the DMI coefficient is defined with respect to the outward normal of surfaces versus the $z$ direction (positive if parallel and negative if anti-parallel.) The results are reported in \figref{fig:DMI}. The flower state is enhanced at the surface where current is injected, \ie, for DMI $> 0$ (case (b) in \figref{fig:DMI}), while it is instead decreased and even inverted when DMI $< 0$ (case (c) in \figref{fig:DMI}). Surprisingly at first sight, there is also an impact at the surface opposite to the one where the DMI is applied. We believe that this occurs because of the high value of the DMI coefficient considered and the height of the nanopillar being only a few times the dipolar exchange length, so that both interfaces are coupled via exchange. The magnitude of the effect at the opposite interface is dramatic, the flower state being suppressed or greatly enhanced depending on the sign of DMI at the other interface. As we have shown that the flower state plays a direct role in the incubation time and dynamical mechanism, in a practical situation with a given material stack it appears crucial to consider whether DMI occurs at the interfaces.

 To conclude, iPMA and DMI may decrease or enhance the flower state. We expected that this should have a direct impact on the incubation time.

\bibliography{main.bib}

\begin{thebibliography}{10}

\bibitem{FER2017}
Amalio Fernández-Pacheco, Robert Streubel, Olivier Fruchart, Riccardo Hertel,
  Peter Fischer, and Russell~P. Cowburn.
\newblock Three-dimensional nanomagnetism.
\newblock {\em Nature Communications}, 8(1), June 2017.

\bibitem{STR2021}
Robert Streubel, Evgeny~Y. Tsymbal, and Peter Fischer.
\newblock Magnetism in curved geometries.
\newblock {\em Journal of Applied Physics}, 129(21), June 2021.

\bibitem{ONO1999}
T.~Ono, H.~Miyajima, K.~Shigeto, K.~Mibu, N.~Hosoito, and T.~Shinjo.
\newblock Propagation of a magnetic domain wall in a submicrometer magnetic
  wire.
\newblock {\em Science}, 284(5413):468--470, April 1999.

\bibitem{CHU2015}
A.~V. Chumak, V. I. Vasyuchka, A. A. Serga, and B.~Hillebrands.
\newblock Magnon spintronics.
\newblock {\em Nature Physics}, 11(6):453--461, June 2015.

\bibitem{RAF2022}
David Raftrey, Aurelio Hierro-Rodriguez, Amalio Fernandez-Pacheco, and Peter
  Fischer.
\newblock The road to 3-dim nanomagnetism: Steep curves and architectured
  crosswalks.
\newblock {\em Journal of Magnetism and Magnetic Materials}, 563:169899,
  December 2022.

\bibitem{BUR2020}
Edward~C. Burks, Dustin~A. Gilbert, Peyton~D. Murray, Chad Flores, Thomas~E.
  Felter, Supakit Charnvanichborikarn, Sergei~O. Kucheyev, Jeffrey~D. Colvin,
  Gen Yin, and Kai Liu.
\newblock 3d nanomagnetism in low density interconnected nanowire networks.
\newblock {\em Nano Letters}, 21(1):716--722, December 2020.

\bibitem{MEN2021}
Fanfan Meng, Claire Donnelly, Luka Skoric, Aurelio Hierro-Rodriguez, Jung-wei
  Liao, and Amalio Fernández-Pacheco.
\newblock Fabrication of a 3d nanomagnetic circuit with multi-layered materials
  for applications in spintronics.
\newblock {\em Micromachines}, 12(8):859, July 2021.

\bibitem{PER2018}
N.~Perrissin, S.~Lequeux, N.~Strelkov, A.~Chavent, L.~Vila, L.~D.
  Buda-Prejbeanu, S.~Auffret, R.~C. Sousa, I.~L. Prejbeanu, and B.~Dieny.
\newblock A highly thermally stable sub-20 nm magnetic random-access memory
  based on perpendicular shape anisotropy.
\newblock {\em Nanoscale}, 10(25):12187--12195, 2018.

\bibitem{WAT2018}
K.~Watanabe, B.~Jinnai, S.~Fukami, H.~Sato, and H.~Ohno.
\newblock Shape anisotropy revisited in single-digit nanometer magnetic tunnel
  junctions.
\newblock {\em Nature Communications}, 9(1), February 2018.

\bibitem{IGA2024}
Junta Igarashi, Butsurin Jinnai, Kyota Watanabe, Takanobu Shinoda, Takuya
  Funatsu, Hideo Sato, Shunsuke Fukami, and Hideo Ohno.
\newblock Single-nanometer cofeb/mgo magnetic tunnel junctions with
  high-retention and high-speed capabilities.
\newblock {\em npj Spintronics}, 2(1), January 2024.

\bibitem{CAC2021}
N.~Caçoilo, S.~Lequeux, B.M.S. Teixeira, B.~Dieny, R.C. Sousa, N.A. Sobolev,
  O.~Fruchart, I.L. Prejbeanu, and L.D. Buda-Prejbeanu.
\newblock Spin-torque-triggered magnetization reversal in magnetic tunnel
  junctions with perpendicular shape anisotropy.
\newblock {\em Physical Review Applied}, 16(2):024020, August 2021.

\bibitem{TIM2015}
A.~A. Timopheev, R.~Sousa, M.~Chshiev, L.~D. Buda-Prejbeanu, and B.~Dieny.
\newblock Respective influence of in-plane and out-of-plane spin-transfer
  torques in magnetization switching of perpendicular magnetic tunnel
  junctions.
\newblock {\em Physical Review B}, 92(10):104430, September 2015.

\bibitem{feeLLGood}
Jean-Chritophe Toussaint.
\newblock feellgood – an fem micromagnetic simulator, 2025.
\newblock Accessed: 2025-02-26.

\bibitem{MAT1997}
Claudio Mattiussi.
\newblock An analysis of finite volume, finite element, and finite difference
  methods using some concepts from algebraic topology.
\newblock {\em Journal of Computational Physics}, 133(2):289--309, May 1997.

\bibitem{FAL2000}
N.A. Fallah, C.~Bailey, M.~Cross, and G.A. Taylor.
\newblock Comparison of finite element and finite volume methods application in
  geometrically nonlinear stress analysis.
\newblock {\em Applied Mathematical Modelling}, 24(7):439--455, June 2000.

\bibitem{GEU2009}
Christophe Geuzaine and Jean‐François Remacle.
\newblock Gmsh: A 3‐d finite element mesh generator with built‐in pre‐
  and post‐processing facilities.
\newblock {\em International Journal for Numerical Methods in Engineering},
  79(11):1309--1331, May 2009.

\bibitem{SI2010}
H.~Si and K.~Gärtner.
\newblock 3d boundary recovery by constrained delaunay tetrahedralization.
\newblock {\em International Journal for Numerical Methods in Engineering},
  85(11):1341--1364, September 2010.

\bibitem{REB1993}
S.~Rebay.
\newblock Efficient unstructured mesh generation by means of delaunay
  triangulation and bowyer-watson algorithm.
\newblock {\em Journal of Computational Physics}, 106(1):125--138, May 1993.

\bibitem{SZA_Nov_2008}
H.~Szambolics, J.-C. Toussaint, L.D. Buda-Prejbeanu, F.~Alouges, E.~Kritsikis,
  and O.~Fruchart.
\newblock Innovative weak formulation for the landau–lifshitz–gilbert
  equations.
\newblock {\em IEEE Transactions on Magnetics}, 44(11):3153--3156, November
  2008.

\bibitem{KRI2014}
E.~Kritsikis, A.~Vaysset, L.D. Buda-Prejbeanu, F.~Alouges, and J.-C. Toussaint.
\newblock Beyond first-order finite element schemes in micromagnetics.
\newblock {\em Journal of Computational Physics}, 256:357--366, January 2014.

\bibitem{ALO2014}
François Alouges, Evaggelos Kritsikis, Jutta Steiner, and Jean-Christophe
  Toussaint.
\newblock A convergent and precise finite element scheme for
  landau–lifschitz–gilbert equation.
\newblock {\em Numerische Mathematik}, 128(3):407--430, February 2014.

\bibitem{SLO1996}
J.C. Slonczewski.
\newblock Current-driven excitation of magnetic multilayers.
\newblock {\em Journal of Magnetism and Magnetic Materials},
  159(1–2):L1–L7, June 1996.

\bibitem{ZHA2004}
S.~Zhang and Z.~Li.
\newblock Roles of nonequilibrium conduction electrons on the magnetization
  dynamics of ferromagnets.
\newblock {\em Physical Review Letters}, 93(12):127204, September 2004.

\bibitem{BUT2001}
W.~H. Butler, X.-G. Zhang, T.~C. Schulthess, and J.~M. MacLaren.
\newblock Spin-dependent tunneling conductance of fe mgo fe sandwiches.
\newblock {\em Physical Review B}, 63(5):054416, January 2001.

\bibitem{YUA2007}
S~Yuasa and D~D Djayaprawira.
\newblock Giant tunnel magnetoresistance in magnetic tunnel junctions with a
  crystalline mgo(001) barrier.
\newblock {\em Journal of Physics D: Applied Physics}, 40(21):R337–R354,
  October 2007.

\bibitem{KAL2009}
Alan Kalitsov, Mairbek Chshiev, Ioannis Theodonis, Nicholas Kioussis, and W.~H.
  Butler.
\newblock Spin-transfer torque in magnetic tunnel junctions.
\newblock {\em Physical Review B}, 79(17):174416, May 2009.

\bibitem{STU2016}
M.~Sturma, C.~Bellegarde, J.-C. Toussaint, and D.~Gusakova.
\newblock Simultaneous resolution of the micromagnetic and spin transport
  equations applied to current-induced domain wall dynamics.
\newblock {\em Physical Review B}, 94(10):104405, September 2016.

\bibitem{MOO1999}
Jagadeesh~S. Moodera and George Mathon.
\newblock Spin polarized tunneling in ferromagnetic junctions.
\newblock {\em Journal of Magnetism and Magnetic Materials},
  200(1–3):248--273, October 1999.

\bibitem{TAO2024}
Ang Tao, Yixiao Jiang, Shanshan Chen, Yuqiao Zhang, Yi~Cao, Tingting Yao,
  Chunlin Chen, Hengqiang Ye, and Xiu-Liang Ma.
\newblock Ferroelectric polarization and magnetic structure at domain walls in
  a multiferroic film.
\newblock {\em Nature Communications}, 15(1), July 2024.

\bibitem{VAL1993}
T.~Valet and A.~Fert.
\newblock Theory of the perpendicular magnetoresistance in magnetic
  multilayers.
\newblock {\em Physical Review B}, 48(10):7099--7113, September 1993.

\bibitem{ARR1979}
A.~S. Arrott, B.~Heinrich, T.~L. Templeton, and Amikam Aharoni.
\newblock Micromagnetics of curling configurations in magnetically soft
  cylinders.
\newblock {\em Journal of Applied Physics}, 50(B3):2387--2389, March 1979.

\bibitem{SAT1989}
M.~Sato and Y.~Ishii.
\newblock Simple and approximate expressions of demagnetizing factors of
  uniformly magnetized rectangular rod and cylinder.
\newblock {\em Journal of Applied Physics}, 66(2):983--985, July 1989.

\bibitem{SCH1988}
Manfred~E. Schabes and H.~Neal Bertram.
\newblock Magnetization processes in ferromagnetic cubes.
\newblock {\em Journal of Applied Physics}, 64(3):1347--1357, August 1988.

\bibitem{RAV1998}
W~Rave, K~Ramstöck, and A~Hubert.
\newblock Corners and nucleation in micromagnetics.
\newblock {\em Journal of Magnetism and Magnetic Materials}, 183(3):329--333,
  March 1998.

\bibitem{SLA2010}
Valeriy~V. Slastikov.
\newblock A note on configurational anisotropy.
\newblock {\em Proceedings of the Royal Society A: Mathematical, Physical and
  Engineering Sciences}, 466(2123):3167--3179, April 2010.

\bibitem{THI2005}
A~Thiaville, Y~Nakatani, J~Miltat, and Y~Suzuki.
\newblock Micromagnetic understanding of current-driven domain wall motion in
  patterned nanowires.
\newblock {\em Europhysics Letters (EPL)}, 69(6):990--996, March 2005.

\bibitem{SUN2000}
J.~Z. Sun.
\newblock Spin-current interaction with a monodomain magnetic body: A model
  study.
\newblock {\em Physical Review B}, 62(1):570--578, July 2000.

\bibitem{BED2010}
D.~Bedau, H.~Liu, J.-J. Bouzaglou, A.~D. Kent, J.~Z. Sun, J.~A. Katine, E.~E.
  Fullerton, and S.~Mangin.
\newblock Ultrafast spin-transfer switching in spin valve nanopillars with
  perpendicular anisotropy.
\newblock {\em Applied Physics Letters}, 96(2), January 2010.

\bibitem{TAK2015}
Y.~Takeuchi, H.~Sato, S.~Fukami, F.~Matsukura, and H.~Ohno.
\newblock Temperature dependence of energy barrier in cofeb-mgo magnetic tunnel
  junctions with perpendicular easy axis.
\newblock {\em Applied Physics Letters}, 107(15), October 2015.

\bibitem{SUPP2025}
See supplemental material for the animations of magnetisation reversal for the
  current velues reported in fig. 6.

\bibitem{KIT1948}
Charles Kittel.
\newblock On the theory of ferromagnetic resonance absorption.
\newblock {\em Physical Review}, 73(2):155--161, January 1948.

\bibitem{ARR1979b}
A.~Arrott, B.~Heinrich, and A.~Aharoni.
\newblock Point singularities and magnetization reversal in ideally soft
  ferromagnetic cylinders.
\newblock {\em IEEE Transactions on Magnetics}, 15(5):1228--1235, September
  1979.

\bibitem{THI2003}
André Thiaville, José~Miguel García, Rok Dittrich, Jacques Miltat, and
  Thomas Schrefl.
\newblock Micromagnetic study of bloch-point-mediated vortex core reversal.
\newblock {\em Physical Review B}, 67(9):094410, March 2003.

\bibitem{RUI2017}
A.~Ruiz-Calaforra, A.~Purbawati, T.~Brächer, J.~Hem, C.~Murapaka, E.~Jiménez,
  D.~Mauri, A.~Zeltser, J.~A. Katine, M.-C. Cyrille, L.~D. Buda-Prejbeanu, and
  U.~Ebels.
\newblock Frequency shift keying by current modulation in a mtj-based stno with
  high data rate.
\newblock {\em Applied Physics Letters}, 111(8), August 2017.

\bibitem{CAC2023(Cacoilo-PhD-PSA)}
Nuno Caçoilo.
\newblock {\em Fundamentals of ultra scaled 3D magnetic tunnel junctions
  andintegration routes for high density memory arrays}.
\newblock PhD thesis, Université Grenoble Alpes, 2023.

\bibitem{EVA2015}
R.~F.~L. Evans, U.~Atxitia, and R.~W. Chantrell.
\newblock Quantitative simulation of temperature-dependent magnetization
  dynamics and equilibrium properties of elemental ferromagnets.
\newblock {\em Physical Review B}, 91(14):144425, April 2015.

\bibitem{WOO2015}
C.~H. Woo, Haohua Wen, A.~A. Semenov, S.~L. Dudarev, and Pui-Wai Ma.
\newblock Quantum heat bath for spin-lattice dynamics.
\newblock {\em Physical Review B}, 91(10):104306, March 2015.

\bibitem{CHE2021}
Runze Chen, Xinran Wang, Houyi Cheng, Kyu-Joon Lee, Danrong Xiong, Jun-Young
  Kim, Sai Li, Hongxin Yang, Hongchao Zhang, Kaihua Cao, Mathias Kläui,
  Shouzhong Peng, Xueying Zhang, and Weisheng Zhao.
\newblock Large dzyaloshinskii-moriya interaction and room-temperature
  nanoscale skyrmions in cofeb/mgo heterostructures.
\newblock {\em Cell Reports Physical Science}, 2(11):100618, November 2021.

\bibitem{GUE2025}
Capucine Gueneau, Fatima Ibrahim, Johanna Fischer, Libor Vojáček,
  Charles-Élie Fillion, Stefania Pizzini, Laurent Ranno, Isabelle Joumard,
  Stéphane Auffret, Jérôme Faure-Vincent, Claire Baraduc, Mairbek Chshiev,
  and Hélène Béa.
\newblock Dzyaloshinskii-moriya interaction chirality reversal with
  ferromagnetic thickness.
\newblock 2025.

\end{thebibliography}

\end{document}